\documentclass{article}
\usepackage{spconf,amsmath,graphicx,cite}
\usepackage{enumitem, comment}
\usepackage{slashbox}
\usepackage{balance}

\title{Black-box Attacks on Automatic Speaker Verification using Feedback-controlled Voice Conversion}
\name{Xiaohai Tian$^1$, Rohan Kumar Das$^1$ and Haizhou Li$^{1,2}$\thanks{This research work is supported by Programmatic Grant No. A1687b0033 from the Singapore Government's Research, Innovation and Enterprise 2020 plan (Advanced Manufacturing and Engineering domain).}}
\address{$^1$Department of Electrical and Computer Engineering, National University of Singapore, Singapore\\
$^2$Kriston AI Lab, China}
\begin{document}
%
\maketitle
\begin{abstract}

Automatic speaker verification (ASV) systems in practice are greatly vulnerable to spoofing attacks. The latest voice conversion technologies are able to produce perceptually natural sounding speech that mimics any target speakers. However, the perceptual closeness to a speaker's identity may not be enough to deceive an ASV system.  In this work, we propose a framework that uses the output scores of an ASV system as the feedback to a voice conversion system. The attacker framework is a black-box adversary that steals one’s voice identity, because it does not require any knowledge about the ASV system but the system outputs. Experimental results conducted on ASVspoof 2019 database confirm that the proposed feedback-controlled voice conversion framework produces adversarial samples that are more deceptive than the straightforward voice conversion, thereby boosting the impostor ASV scores. Further, the perceptual evaluation studies reveal that converted speech do not adversely affect the voice quality from the baseline system.


\end{abstract}
\section{Introduction}
\label{sec:intro}



Automatic Speaker Verification (ASV) systems have enabled many real-world applications~\cite{tomi,sv_debut,smart_home,Das2016,rkd_APSIPA2018_realism}. Such systems for are facing increasing threats from spoofing attacks for unauthorized access~\cite{Li2016_spoof_TD,yamagishi2012}. In general, such spoofing attacks are grouped under four major categories, namely, impersonation, replay, Voice Conversion (VC) and Text-To-Speech (TTS) synthesis~\cite{spoof_review}. To safeguard the ASV systems from spoofing attacks, anti-spoofing countermeasures have become critically important~\cite{spoof_review, ASVspoof_journal}. 


From the the attackers' point of view, the most effective way to perform attack is within the ASV system~\cite{spoof_review}. However, it generally requires system level access, which is not easily possible in practice. Alternatively, the spoofing attacks can performed at the input signal level by VC or TTS. The current VC and TTS systems are generally developed with the objective of improving the perceptual quality and similarity, which makes the generated speech to sound closer to the target~\cite{Kinnunen2018_vc,Obama2018}. However, perceptual similarity may not be enough to deceive ASV systems as they don't perceive in the same way as humans~\cite{hansen_review}.  

The black-box attacks represent an advance level of spoofing attack using synthetic speech~\cite{blackbox_arxiv,blackbox_adversary,adversarial_attacks}. In such attacks, the attacker does not require the knowledge of either internal functionality of a system or training data. The attacker rather needs to observe the output of black-box for a given input~\cite{blackbox_adversary}. The strategy of black-box attacks deals with training a substitute model using machine learning techniques to evade the target system by considering the outputs of black-box. Such attacks were evaluated on computer vision tasks~\cite{blackbox_adversary}. We believe they have become the spoofing adversary to ASV systems.    

Recent studies~\cite{Tomi_attack_mimic,VESTMAN_mimic_ASV} show that ASV system can be exploited to perform assist voice mimicry attacks. The results however showed that even there exists a similarity between voice of an attacker and the target speaker, mimicks still can not attack the ASV system successfully. It is also found from their studies that untrained impersonators do not show a high threat to an ASV system, but the use of an ASV system to attack another ASV system is a potential threat. To bring the previous studies a step forward, we consider training a VC system as the black-box adversary to attack the ASV systems.

In this work, we propose a feedback-control VC system based on Phonetic PosterioGram (PPG) VC approach to perform attacks to an ASV system. It assumes that we don't have any system level access to the ASV system but the ASV output scores. Then the VC network is trained with feedback from the output score of the black-box ASV system. The objective function of this substitute network is modified to have a joint cost based on the VC and the black-box ASV output. Such feedback-controlled VC can steal one's voice identity by simply observing the output scores of the ASV system.
To the best of our knowledge, black-box attacks on ASV systems through feedback-controlled VC have not been investigated yet. The framework of performing black-box attacks on ASV highlights the novelty of the current work.

\section{PPG-based Voice Conversion Framework}
\label{se:PPG_VC}

Voice conversion aims to modify the source speaker's voice to sound like that of the target speaker without changing the linguistic content. Such techniques can be used to generate the voice of any target speakers to attack ASV systems.
Generally, a conversion function is trained with parallel data from source and target speakers~\cite{stylianou1998continuous, toda2007voice, erro2010voice, tian2014correlation, tian2017exemplar, wu2014exemplar, ccicsman2017sparse, xie2014sequence, sisman2018voice}.
Recently, various techniques also have been proposed to achieve non-parallel data VC~\cite{erro2010inca, nakashika2016non}. Among non-parallel data attempts, one of the successful approaches is PPG-based VC~\cite{sun2016phonetic, tian2018average, tian2019ppgwavenet}. It models the relationship between the PPG features, that is called the linguistic feature, to the acoustic feature. 
As the PPG feature is considered to be speaker independent, the source speaker information is not required in the training process.
Hence, PPG-based VC technique is suitable for many to one conversion, which relates to the attack scenario, where an unknown attacker pretends to be the target speaker using converted voice.


\subsection{PPG-based Voice Conversion}
\label{ppg_vc}
PPG-based VC framework consists of two stages: training and conversion. 

During training, PPG and aocustic features are first extracted from the target speaker, denoted as $\mathbf{X} = {[ \mathbf{x}_1, \cdots, \mathbf{x}_N ]}$ and $\mathbf{Y} = {[ \mathbf{y}_1, \cdots, \mathbf{y}_N ]}$, respectively.
$N$ denotes the number of feature frame.
As $\mathbf{X}$ and $\mathbf{Y}$ are extracted from the same utterance, these two feature sequence are initially aligned. Then, the feature mapping of the target speaker is trained by a conversion network with a criterion of minimizing the mean squared error (MSE) between target and predicted features. Given a predicted acoustic feature sequence denoted as $\widehat{\mathbf{Y}} = {[ \widehat{\mathbf{y}_1}, \cdots, \widehat{\mathbf{y}_N} ]}$, the MSE can be computed by
\begin{equation}
\label{eq:MSE_VC_train}
\text{Loss}_{VC} = \frac{1}{N} (\mathbf{Y} - \widehat{\mathbf{Y}})^\top (\mathbf{Y} - \widehat{\mathbf{Y}}).
\end{equation}

During conversion, the trained conversion model is used to generate the converted acoustic features for given PPGs from any source speaker. Finally, a vocoder is used to reconstruct the speech signal from the acoustic features.

\subsection{Limitation for ASV Attacks}

The VC performance is typically evaluated in terms of perceptual quality and similarity by human listeners~\cite{sun2016phonetic, tian2018average}.
However, improved perceptual quality may not always correlate with an improved ASV score as far as spoofing attacks are concerned. Such observations have been reported in previous works~\cite{hansen_review, Tomi_attack_mimic}. From the attackers' point of view, a joint objective function that includes the ASV score as the feedback will directly optimize the VC outputs to become more deceptive to the system.


\section{Black-box Attacks on ASV using feedback-controlled VC}
\label{secii}

This section discusses the proposed framework of performing black-box attacks on ASV using feedback-controlled VC. The ASV system acts as a black-box that provides feedback to the VC system to have the desired output. 

\subsection{ASV System as a Black-Box}

The progress in the field of ASV research has witnessed major breakthrough in the recent years. It has evolved a lot from classical GMMs to factor analysis and deep learning models~\cite{tomi,Dehak2011,xvectors}. In this work, we consider a ASV system with i-vector speaker models as a black-box~\cite{Dehak2011}. Although recent studies on x-vectors have gained attention in the community~\cite{xvectors}, the i-vector based system is still popular in the community due to its effectiveness under controlled conditions. An i-vector is a compact representation that projects the GMM mean supervector of an utterance to a lower dimensional space that contains the dominant speaker characteristics. 
We call the ASV system as a black-box because the attacker doesn't have any access to internal algorithms except the output scores.


\subsection{Feedback-controlled VC}

\begin{figure*}[!htb]
\centering
\includegraphics[width=14.5cm]{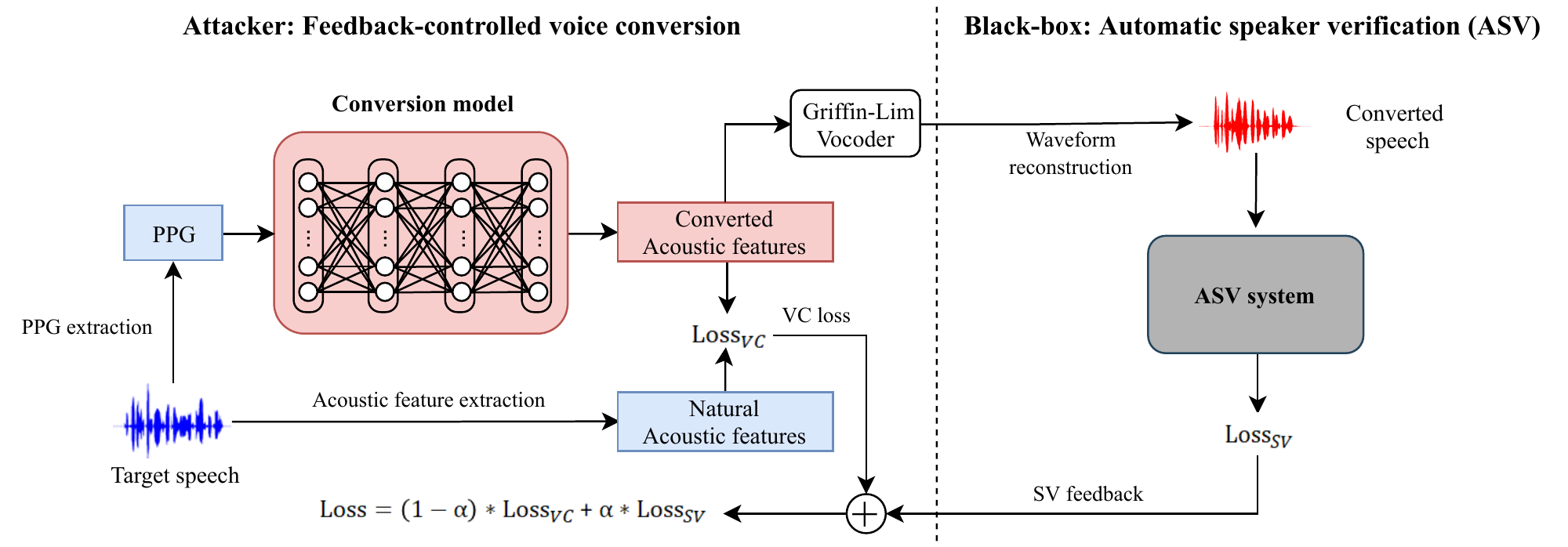}
\caption{The block diagram of the training stage for the feedback-controlled voice conversion and ASV system. $\text{Loss}_{VC}$ and $\text{Loss}_{SV}$ denote the loss from conversion model and ASV feedback respectively. The weighted sum of $\text{Loss}_{VC}$ and $\text{Loss}_{SV}$ is used to update the conversion model.}
\label{fig:VC_SV}
\vspace{-2mm}
\end{figure*}

The objective of the feedback-controlled VC is to adjust the converted voice to be more deceptive for the ASV black-box.
It is assumed that the attacker has the knowledge of the ASV output score for each input trial. However, the attacker does not need to have the understanding of the ASV system.


In our proposed feedback-controlled VC, an ASV feedback is included into the training process to update the VC model.
Fig.~\ref{fig:VC_SV} shows the training process of the feedback-controlled VC system.
Two types of losses are used to update the conversion model by minibatch stochastic gradient descent. 
For each minibatch, the predicted acoustic features, $\widehat{\mathbf{Y}}$, is first obtained with given input PPGs, $\mathbf{X}$, as described in Section~\ref{ppg_vc}.
The MSE loss of VC, $\text{Loss}_{VC}$, can be calculated by Eq.~(\ref{eq:MSE_VC_train}).
Simultaneously, we generate the converted speech with $\widehat{\mathbf{Y}}$.
An ASV score, denoting the similarity between i-vectors of converted and target speech, is obtained by feeding the converted speech to the black-box ASV system.
The $\text{Loss}_{SV}$ is calculated by normalizing it between $[0, 1]$, followed by its transformation in the negative scale as another loss.
Then the two parts of losses are combined with a weighted ratio $\alpha$ to update the conversion network. The combined loss is defined as
\begin{equation}
\label{eq:LOSS_VC_train}
\text{Loss} = (1-\alpha) \times \text{Loss}_{VC} + \alpha \times \text{Loss}_{SV},
\end{equation}

\section{Experiments}
\label{seciii}

This section describes the database and experimental setup. 

\subsection{Database}


The experiments are conducted on the logical access subset of ASVspoof 2019\footnotemark[1] corpus that deals with synthetic speech attacks~\cite{ASVspoof2019_plan}.
For VC system training, 6 speakers form development set are selected as target speakers, including 3 male (LA\_0070, LA\_0071 and LA\_0073) and 3 female (LA\_0069, LA\_0072 and LA\_0074) speakers. 120 utterances from each speaker are used for training, where the size of training and validation sets are fixed as 100 and 20 utterances, respectively. 42-dimensional PPG features are extracted by the PPG extractor~\cite{tian2018average} as VC system input. We then extract 80-dimensional mel-filterbank features from 513-dimensional spectrogram as VC system output. The Griffin-Lim vocoder~\cite{griffin1984signal} is used for speech signal reconstruction. All audio files are sampled at 16 kHz for VC.

For ASV attack experiment, 67 speakers from the evaluation set are selected, with 20 utterances from each speaker. These speakers form the imposter set for our studies totalling 1,340 non-target trials against each target speaker. The same 1,340 utterances are used as source speech for two VC systems with and without ASV feedback. The two sets of converted trials are then considered to perform attacks to the ASV system. In addition, we consider 120 genuine trials from each of the 6 target speakers to obtain the genuine scores. We note that all audio files are resampled at 8 kHz for ASV studies.





\subsection{Experimental Setup}


\begin{itemize}[leftmargin=*,topsep=0pt]
\itemsep0em
    \item \textbf{PPG-VC}: It refers to the PPG-based VC system without ASV feedback~\cite{sun2016phonetic}. The network contains two Bidirectional Long Short-Term Memory (BLSTM) layers with 512 hidden units of each layer. The network input is PPG features (42-dim); while the dimension of output is 240, consisting of the mel-filterbank (80-dim) with its dynamic and accelerate features. 
    \item \textbf{PPG-VC-FC}: It refers to the proposed feedback-controlled VC. We use the same network setting as that of PPG-VC system. The network is updated with the combined loss from both VC and ASV systems.
    Empirically, the weighted ratio $\alpha$ is set to 0.7 according to the observations on the validation set.
\end{itemize}

For the training of all the models, the minibatch size is set to 5, while the momentum and learning rate are set as 0.9 and 0.002, respectively.

We use i-vector based system for ASV studies~\cite{Dehak2011}. The attacker uses this system as a black-box to feed the output score to the proposed feedback-controlled VC. The standard Kaldi\footnotemark[2] implementation of i-vector is used in this work. The system uses mel frequency cepstral coefficient (MFCC) as acoustic features and energy based voice activity detection for feature selection. The Switchboard and NIST SRE corpus 2006-2012 are used to train the i-vector extractor of 400 speaker factors. The extractor is then used to derive 400-dimensional i-vector that represents the utterances of every speaker by dominant speaker representation. In this work, we used cosine distance between the train and the test i-vectors to generate the ASV score for authentication of a trial. We consider equal error rate (EER) that provides the trade-off between the false alarm and miss probability as the metric to report results.   

\footnotetext[1]{https://datashare.is.ed.ac.uk/handle/10283/3336}
\footnotetext[2]{http://kaldi-asr.org/}


\section{Results and Analysis}
\label{seciv}

In this section, we report the results for the black-box attacks performed on the ASV system. Apart from the studies with the attacks, perceptual studies are also conducted. 


\subsection{ASV Studies with Black-box Attacks}



Table~\ref{table1} shows the performance comparison for ASV systems under different attacks, which include imposters, PPG-VC and proposed PPG-VC-FC. The performance are reported for male and female speakers separately as well as the overall result. We observe that the ASV system perform very effectively when the imposter trials are used. However, the performance decreases by large margin when the PPG-VC based VC attacks are performed. It is observed that our proposed PPG-VC-FC based system further decreases the performance showing the vulnerability of ASV systems to black-box attacks. In order to have detailed investigation, we observe the score distributions of different systems with two examples.    


\begin{table} [t!]
\caption{\label{table1} { Performance in EER (\%) for ASV system under different attacks on evaluation set of ASVspoof 2019 corpus.}}
\vspace{2mm}
\centerline{
\begin{tabular}{|c|*{3}{c|}}\hline
\backslashbox{Subset}{Attacks}
&\makebox[3em]{\bf Imposter}&\makebox[3em]{\bf PPG-VC}&\makebox[5em]{\bf PPG-VC-FC}\\
\hline \hline
Male & 2.62 & 25.00 & \textbf{26.67}\\
Female & 2.38 & 31.60 & \textbf{32.90}\\
\hline \hline
Overall & 2.72 & 29.25 & \textbf{30.73}\\
\hline
\end{tabular}}
\vspace{-2mm}
\end{table}



\begin{figure}[!h]
\centering
\includegraphics[width=8.5cm]{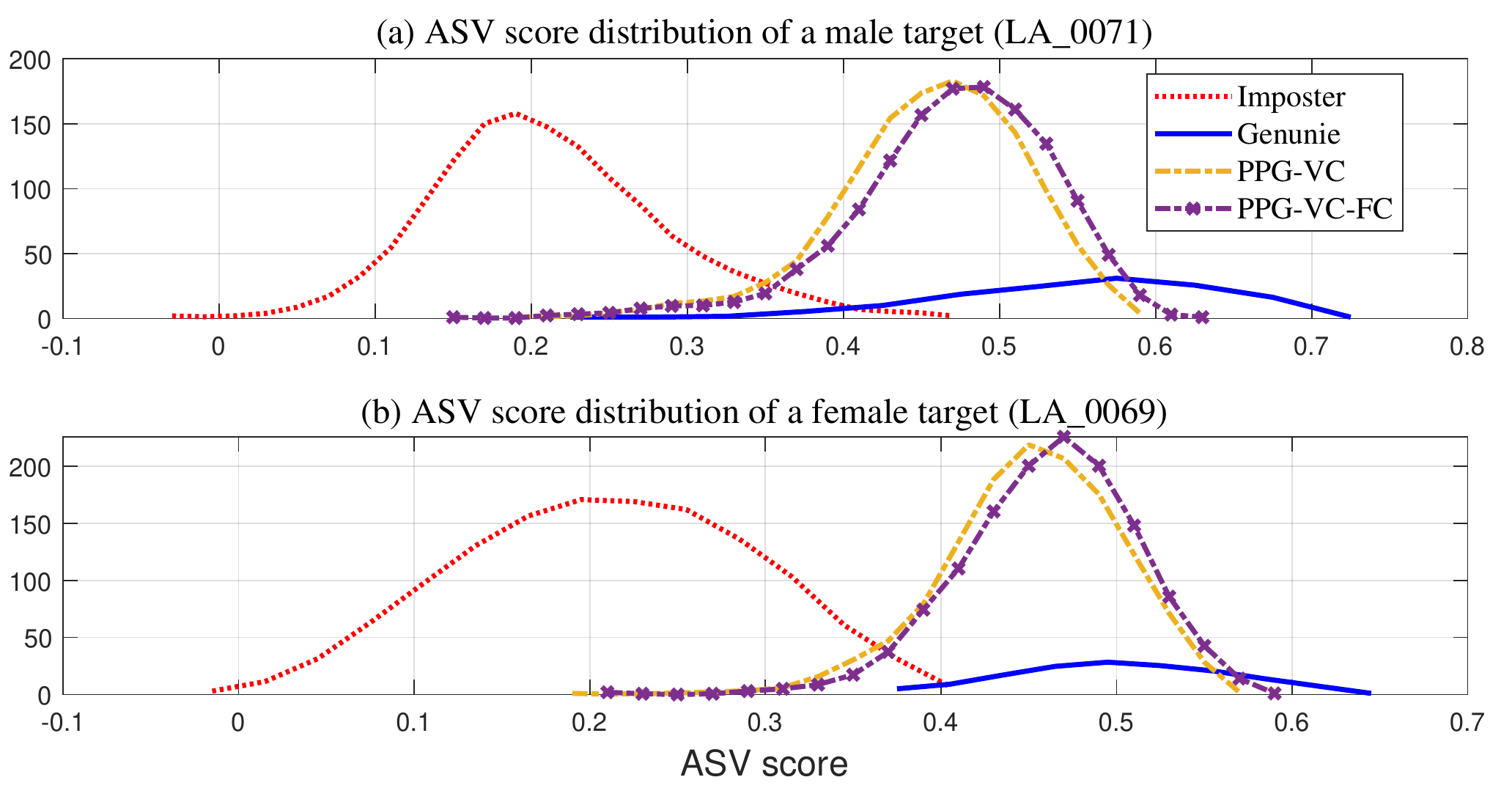}
\caption{ASV score distributions for a male (LA\_0071) and a female (LA\_0069) target speaker. PPG-VC and PPG-VC-FC legends represent scores from baseline VC and our proposed feedback-controlled VC, respectively.}
\label{fig:SV_score}
\end{figure}

Fig.~\ref{fig:SV_score} (a) and (b) show examples of the score distributions for a male (LA\_0071) and a female (LA\_0069) target speaker. We observe that for both male and female speaker, the genuine and imposter scores generated from the ASV system are clearly separated that suggests its effectiveness for verifying a claimed identity. It is found that PPG-VC system can yield ASV scores close to those of the genuine target samples. In addition, we observe that our proposed PPG-VC-FC further enhances the ASV scores by shifting the score distributions to the right. The results clearly show that PPG-VC-FC heightens the security threat to the ASV system. 

In summary, we observe that the PPG-VC-FC system effectively shifts the ASV output scores towards those of the target speakers. While the experiments in this paper are conducted on the back-box ASV system, it is worth studying how the same feedback-controlled mechanism interacts with anti-spoofing countermeasures as well.


\subsection{Perceptual Evaluation}
\begin{figure}[!t]
\centering
\includegraphics[width=8.5cm]{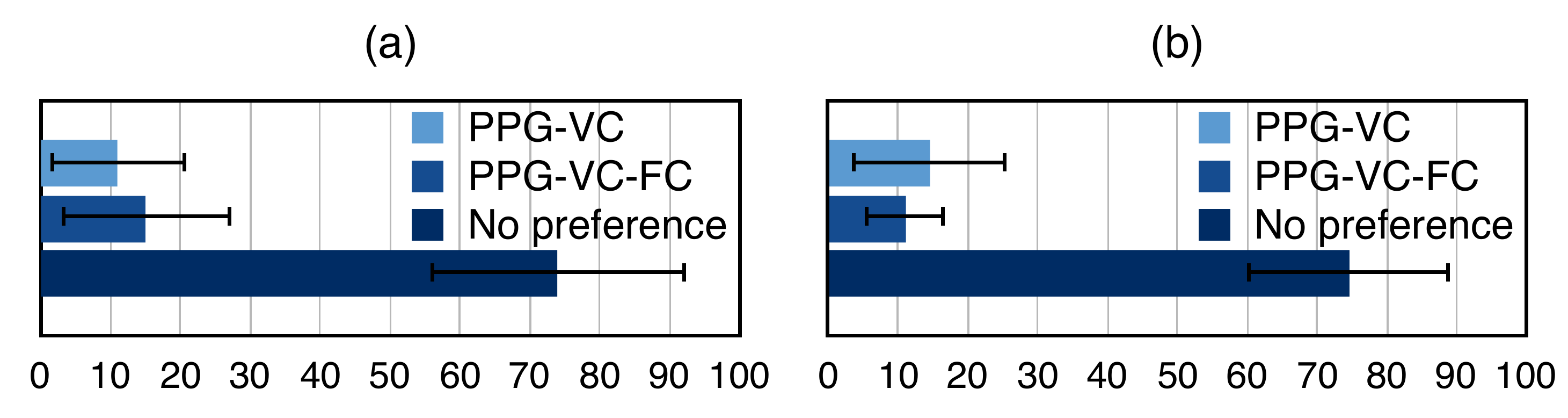}
\vspace{1mm}
\caption{ Perceptual evaluation results in terms of both quality and similarity. (a) Results of quality preference tests with 95\% confidence intervals for PPG-VC and PPG-VC-FC. (b) Results of similarity preference tests with 95\% confidence intervals for PPG-VC and PPG-VC-FC.}
\label{fig:quality}
\end{figure}



We further perform perceptual evaluation on the attacks generated from conventional VC attacks and our proposed black-box. This may provide some insights into how humans perceive the deceptiveness of converted speech.  

The AB and ABX preference tests are conducted to assess the speech quality and speaker similarity respectively. 
During tests, each paired samples A and B are randomly selected from the proposed PPG-VC-FC and baseline PPG-VC, respectively. Each listener is asked to choose the sample with better quality/similarity or no preference. 
For each test, 20 sample pairs are randomly selected from the 1,340 paired samples. We invite 20 subjects to participate in both tests.

Fig.~\ref{fig:quality} (a) and (b) present the results of quality and similarity preference tests with 95\% confidence respectively.
It is observed that for both quality and similarity tests the identification rates of PPG-VC and PPG-VC-FC fall into each other’s confidence intervals. This indicates that they are not significantly different in terms of speech quality and speaker identity.
In addition, most of the listeners could not differentiate between the two. 
It shows that the feedback-controlled VC is able to attack the ASV system while maintaining the speech quality and similarity.

\section{Conclusions}
\label{conc}

In this work, we study black-box attacks on ASV systems with a feedback-controlled VC system. Although the vulnerability traditional spoofing attacks to ASV has been established, the black-box attacks to machine learning poise a greater threat. We use ASV system outputs as the feedback to increase the deceptiveness of the converted voice. The studies conducted on ASVspoof 2019 corpus suggest that the proposed feedback-controlled VC system is able to enhance the ASV scores of the converted speech that makes it more deceptive towards authentication against the target speakers. 
Moreover, according to the subjective test results, we find that feedback-controlled VC maintains the performance in terms of speech quality and speaker similarity for the generated speech.

%
\newpage
\footnotesize
\bibliographystyle{IEEEbib}
\balance
\bibliography{MyReferences_new}

\end{document}